\documentclass[9pt,twocolumn,twoside]{opticajnl}
\journal{opticajournal}

\setboolean{shortarticle}{true}

\title{Coherence analysis of tightly locked mid-infrared quantum cascade laser frequency combs}

\author[1,*]{Alexandre~Parriaux}
\author[1]{Kenichi~N.~Komagata}
\author[2]{Mathieu~Bertrand}
\author[1]{Valentin~J.~Wittwer}
\author[2]{Jérôme~Faist}
\author[1]{Thomas~S\"udmeyer}

\affil[1]{Laboratoire Temps-Fréquence, Institut de Physique, Université de Neuchâtel, Avenue de Bellevaux 51, 2000 Neuchâtel, Switzerland}
\affil[2]{Institute for Quantum Electronics, ETH Zurich, Auguste-Piccard-Hof 1, 8093 Zurich, Switzerland}

\affil[*]{alexandre.parriaux@unine.ch}

\begin{abstract}
Frequency combs are powerful tools for many applications and high performances are achieved by stabilizing these lasers.
For operation in the mid-infrared, quantum cascade lasers (QCL) are ideal candidates as they present numerous advantages.
However, stabilized QCL-combs lack of a detailed characterization of their noise properties due to the sensitivity limits of current analyzing techniques.
To overcome these challenges, we developed what we believe to be the first tightly locked dual QCL-comb system. Its use is twofold.
First, phase noise analysis of the dual-comb signal shows residual phase noise below 600~mrad for all comb lines, and we characterize the comb coherence as well as the performances of the repetition frequency locking mechanism.
Second, we demonstrate coherent averaging with a $7\times 10^5$~Hz\textsuperscript{1/2} figure-of-merit system, which is compatible with high precision spectroscopy.
\end{abstract}

\setboolean{displaycopyright}{false} 

\begin{document}

\maketitle

Quantum cascade lasers (QCLs) are semiconductor sources able to emit frequency combs in the mid-infrared (MIR) spectral range~\cite{Hugi-nature-2012,Faist-nanophot-2016}. Compared to other MIR comb sources, they have the advantage of high power, high repetition rate, small footprint, and simple operation. Therefore, QCLs are an effective technology for highly sensitive and selective molecular sensing by means of MIR spectroscopy~\cite{Picque-natphot-2019,komagata2023absolute}.
However, free-running QCLs exhibit typical linewidths around half a MHz~\cite{Cappelli-optica-2015,shehzad2020frequency}, which limits applications requiring high spectral purity, hence restricting the impact and use of QCL-combs without a proper stabilization scheme.
For example, cavity enhanced spectroscopy~\cite{foltynowicz2013cavityenhanced} with QCL-combs could boost the sensitivity to unprecedented levels. QCLs have nevertheless been effective and productive with scientific results through dual-comb spectroscopy (DCS) with computational corrections~\cite{burghoff-sciadv-2016,klocke2018singleshot,sterczewski2019computational}. 
Also there, tight stabilization could lead to higher sensitivity through continuous and long-term coherent averaging~\cite{Picque-natphot-2019}.

While the full stabilization of MIR QCL-combs has already been reported~\cite{komagata2022absolute}, several aspects still need to be addressed. First, the actuation of the drive current allows the stabilization of one degree of freedom~\cite{cappelli2016frequency}, but as it exploits the same thermal effects as the noise process~\cite{tombez2012temperature,tombez2013wavelength}, the spectral purity is limited. We recently solved this problem using a faster actuator based on near-infrared (NIR) light illumination of the QCL front facet~\cite{Komagata-aplphot-2023}, but it has not yet been implemented for DCS.
The remaining degree of freedom can be locked by radio frequency (RF) injection locking~\cite{st-jean2014injection}. The coherence of this technique was shown qualitatively by a near-unity intermodal coherence~\cite{burghoff2015evaluating} measured by shifted wave interference Fourier transform (SWIFT) spectroscopy~\cite{Hillbrand-natphot-2019}. However, a quantitative characterization of the conformity of the repetition rate to the master oscillator is lacking.
In fact, preliminary measurements suggested the presence of phase noise greater than the already inconvenient synthesizer noise multiplicative accumulation~\cite{consolino2019fully,Komagata-oe-2021}.

In this letter we report for the first time, to the best of our knowledge, a stabilization scheme for the full and effective tight-lock of a dual QCL-comb system. Such a performance is achieved thanks to the NIR light actuator that enhances the low efficiency of current actuation. 
A thorough analysis of the RF multi-heterodyne beating shows that all comb lines are tightly phase-locked with less than 600~mrad of residual phase noise with 1-s integration time.
Moreover, the obtained results are also used to characterize the performance of RF injection locking with an unprecedented sensitivity, and we demonstrate its limits.
Finally, we show that our stabilized dual-comb setup can be used for coherent averaging, paving the way towards high precision DCS.

\begin{figure}[t]
\centering
\includegraphics[width=\linewidth]{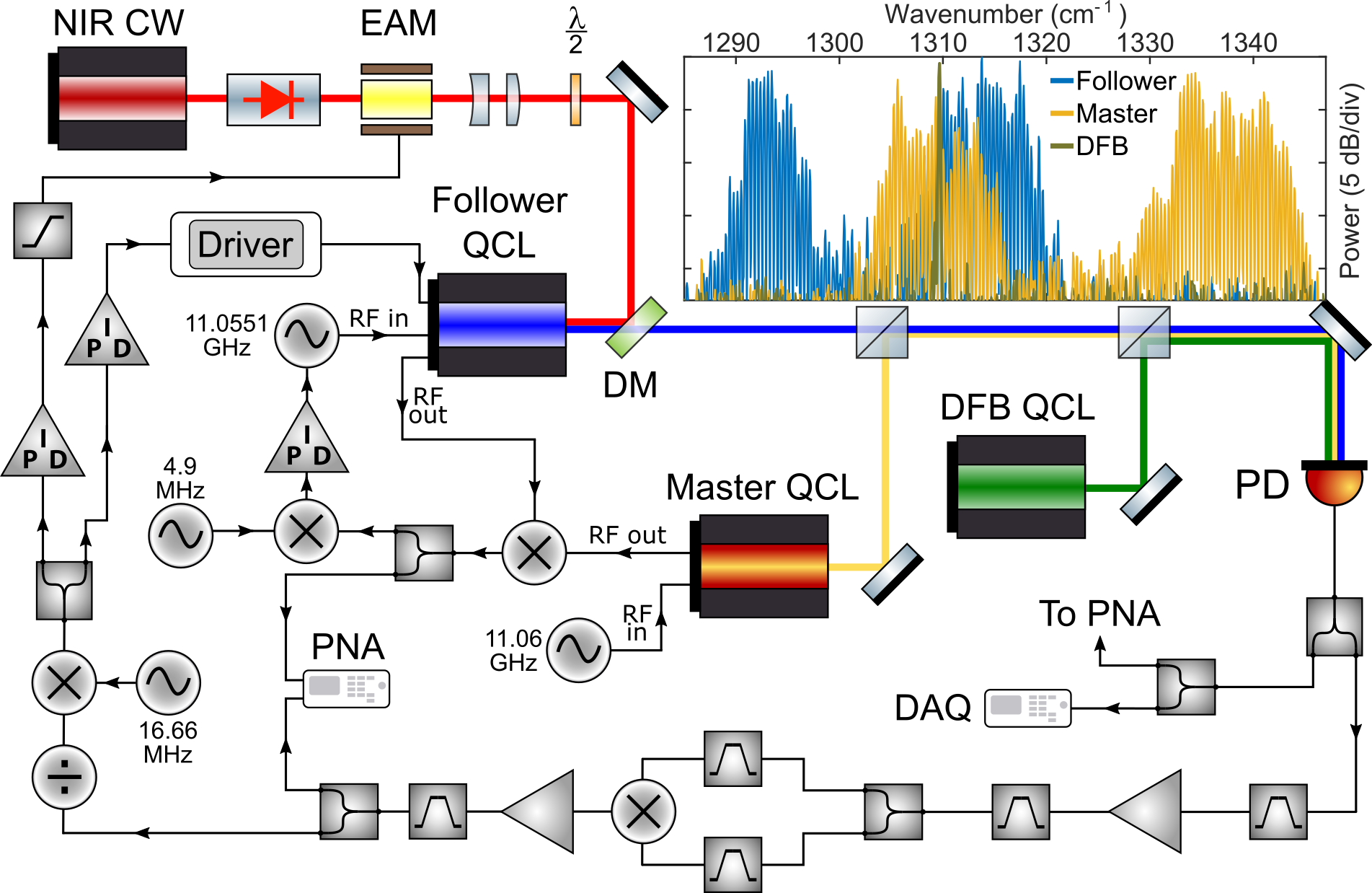}
\caption{Experimental setup used to stabilize one comb line of the dual-comb spectrum by actuation on the follower QCL, and the dual-comb system. The inset shows the optical spectra of the different QCLs. EAM: electro-absorption modulator, DM: dichroic mirror, PD: Photodetector, DAQ: data acquisition system, PNA: phase noise analyzer.}
\label{fig:setup}
\end{figure}

The experimental setup considered here is schematized in \autoref{fig:setup} and is built upon the one we introduced in Ref.~\cite{Komagata-aplphot-2023} for a single frequency comb. We begin with two QCLs combs in a master-follower configuration. Both QCLs are controlled in current and temperature using home made drivers that respectively set the operation points for the master and follower QCL to 1015~mA~/~2\textdegree C and 1200~mA~/~1.2\textdegree C. This leads to two QCL-combs that overlap around 1312~cm$^{-1}$ over more than 70 lines. The inset in \autoref{fig:setup} shows the optical spectra of the different QCLs used here.
The repetition frequencies of the follower and master combs are stabilized by the injection of two 10~dBm signals issued by synthesizers (Agilent~E8257D and R\&S~SMP02) respectively as $f_\text{rep}^\text{FO} = 11.0551$~GHz and $f_\text{rep}^\text{MA} = f_\text{rep}^\text{FO} + \Delta f_\text{rep} $ with $ \Delta f_\text{rep} = 4.9 $~MHz.
For better performance, we also stabilize the frequency difference between both synthesizers against a low noise signal source at $ \Delta f_\text{rep} $. The phase-error signal is fed to the generator locking the follower QCL for phase modulation. Also note that all signal generators are referenced to a maser.

In parallel, a 1.55~µm continuous wave (CW) laser delivering around 1~mW output power is modulated using an integrated electro-absorption modulator. After passing through a half-wave plate, the NIR beam reaches a custom-made dichroic mirror produced by ion beam sputtering that is placed in the path of the follower QCL-comb to illuminate the front facet of the QCL-chip. 
The QCL-comb beams are combined together as well as with a CW distributed feedback (DFB) QCL. The latter is driven at a current of 187~mA and a temperature of 2\textdegree C, to emit at 1309.7~cm$^{-1}$. In this configuration, the frequency difference between the DFB and the nearest comb line is around 670~MHz for the master QCL and around 920~MHz for the follower QCL.

\begin{figure}[t]
\centering
\includegraphics[width=\linewidth]{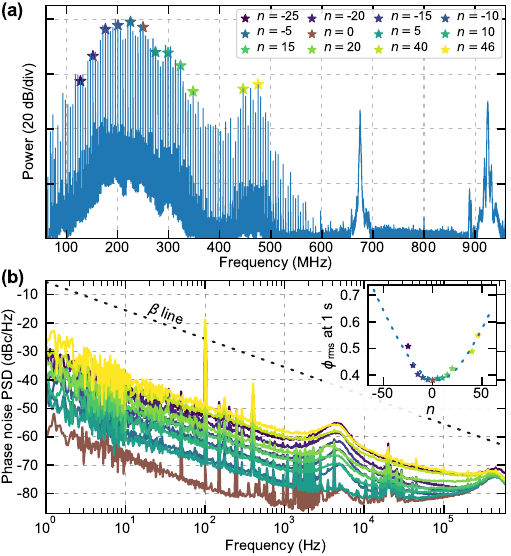}
\caption{\textbf{(a)} RF spectrum showing the dual-comb signal and the beating between the DFB-QCL with one comb line of each QCL-comb. The comb lines are numbered with respect to line 0 that we stabilized. \textbf{(b)} PN-PSD of different comb lines from the dual-comb spectrum presented in \textbf{(a)}. The inset shows the residual phase noise $\phi_\text{rms}$ at 1~s with respect to the line number $n$, and a comparison with a fitted hyperbolic function.}
\label{fig:DC_RF_PN}
\end{figure}

The dual-comb signal and the beating between both combs and the DFB-QCL is captured on a fast photodetector, and a typical spectrum is presented in \autoref{fig:DC_RF_PN}~\textbf{(a)}. Using a set of power splitters, filters and amplifiers, we extract the RF signals associated with the beating between the DFB-QCL and the nearest comb line coming from each QCL-comb, and we mix them together. The resulting signal at frequency $f^\text{RF}_{0} $~=~249.9~MHz corresponds to a particular line of the dual-comb spectrum that we label as line 0.
We then divided it by 15 and mixed it with a synthesizer at 16.66~MHz, resulting in a error signal that is fed to two PID servo-controllers. As in our previous study~\cite{Komagata-aplphot-2023}, one PID acts on the electrical current of the driver of the follower QCL-comb, and the second PID acts on the integrated electro-absorption modulator of the NIR laser for intensity modulation.

Once the stabilization is active, we measured the phase noise power spectral density (PN-PSD) of $ f^\text{RF}_{0} $ using a phase noise analyzer (R\&S~FSWP26). The results show an integrated phase noise between 1~Hz and 10~MHz below 400~mrad, which is twice what we obtained in Ref.~\cite{Komagata-aplphot-2023}. This difference is explained by a longer RF chain and a weaker signal-to-noise ratio.

We now take an interest in the PN-PSD of several lines of the dual-comb spectrum and especially their mutual coherence. Such line-by-line PN-PSD analysis is a possible way to assess the performance of the stabilization scheme, which can easily be performed here on the dual-comb spectrum. The results are shown in \autoref{fig:DC_RF_PN}~\textbf{(b)} where PN-PSD measurements of the comb lines highlighted in \autoref{fig:DC_RF_PN}~\textbf{(a)} are presented, along with the extracted integrated phase noise $\phi_\text{rms}$ at 1-s integration time in inset. First, we can observe that all lines considered are below the $\beta$ line~\cite{didomenico2010simple}.
Second, as expected for a frequency comb, the phase noise increases with the line number $n$ due to the noise multiplication of the repetition rate, and $\phi_\text{rms}$ presents a hyperbolic trend. The bump near 4~kHz originates from the repetition rate as well (see below).

With this set of phase noise measurements, we can isolate the noise contributions to the individual comb lines. Let us consider the PN-PSD of the $n$-th comb line of the dual-comb spectrum $S_n$, which can be written as:
\begin{equation} \label{eq:pn_sn}
    S_n = S_0 + n^2 S_\text{rep}+ n S_\text{corr} \;,
\end{equation}
with $S_\text{rep}$, $S_0$ and $S_\text{corr}$ respectively the PN-PSD of $\Delta f_\text{rep}$, of $f^\text{RF}_{0} $ (i.e., line 0), and of the correlation between $\Delta f_\text{rep}$ and $f^\text{RF}_{0} $. With the knowledge of $S_n$ for three lines of index 0, $l$ and $k$, we can extract $S_\text{rep}$ using \eqref{eq:pn_sn}, which gives us:
\begin{equation} \label{eq:sn_contrib}
S_\text{rep} = \frac{1}{k-l} \left ( \frac{S_k}{k} - \frac{S_l}{l} + S_0 \left ( \frac{1}{l} - \frac{1}{k} \right ) \right ) \;.
\end{equation}
Note that $S_\text{corr}$ can be extracted in the same way.

\begin{figure}[t]
\centering
\includegraphics[width=\linewidth]{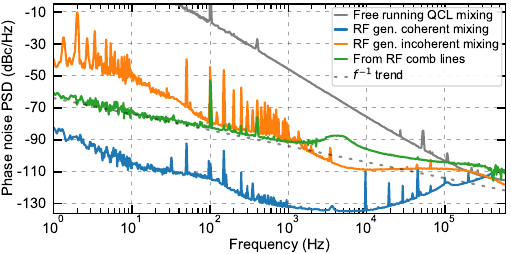}
\caption{PN-PSD of $\Delta f_\text{rep} $ from the mixing of the free-running QCL-comb repetition frequencies, coherent and incoherent mixing of the generators used for RF injection locking in the QCL-combs, and from the analysis of the RF comb lines.}
\label{fig:pn_dfrep}
\end{figure}

\autoref{fig:pn_dfrep} shows the results obtained for $S_\text{rep}$ using \eqref{eq:sn_contrib} with $k=-15$ and $l=20$. These results are compared with the PN-PSD of $\Delta f_\text{rep} $ that we measured electrically by mixing the RF outputs of the QCL-combs (coherent mixing), or by numerically summing the PN-PSD of the generators individually (incoherent mixing). We remind that in our setup the difference of frequency between the generators is stabilized against a third low noise generator, which is why the coherent mixing is much lower than the incoherent mixing. Also note that we could have chosen other lines for the computation of $S_\text{rep}$ as it gives similar results.

Two different noise regimes can be observed for $S_\text{rep}$.
First, below 1~kHz offset frequency, $S_\text{rep}$ has a lower PN-PSD than the incoherent mixing of the generators thanks to the stabilization of $\Delta f_\text{rep} $. Indeed, we verified that when turned off, $S_\text{rep}$ shows no more enhancement compared to the incoherent mixing and matches it (below 1~kHz). However, we can observe that $S_\text{rep}$ is largely above the PN-PSD of the coherent mixing. 
In the second region above 1~kHz offset frequency, although $S_\text{rep}$ is still below the free-running case, the PN-PSD is higher than the incoherent mixing of the signal generators. In particular, a bump near 4~kHz is present which corresponds to the one observed on the PN-PSD of the comb lines in \autoref{fig:DC_RF_PN}~\textbf{(b)}. The bump is reminiscent of the findings in Ref.~\cite{consolino2019fully}.

These results suggest a limit in the stabilization of the repetition frequency via RF injection locking, i.e., the repetition rate could imitate the synthesizer only down to a phase noise limit with a $f^{-1}$ trend (dashed line in \autoref{fig:pn_dfrep}). Above 1~kHz, this limit happens to be above the noise of the 11~GHz synthesizers, which could depend on the injected RF power and frequency. Preliminary measurements were made to verify this assumption by comparing electrically and optically (detected with a very fast photodetector) the PN-PSD of the repetition rate, and we evidenced similar differences. Further investigation will be necessary to draw a full picture of the dynamics and reduce the limit. However, we should note that our scheme already achieves low noise for all comb lines, as shown in \autoref{fig:DC_RF_PN}~\textbf{(b)} and below.

As RF injection locking of MIR QCL-combs has already been thoroughly investigated~\cite{st-jean2014injection,Hillbrand-natphot-2019,kapsalidis2021midinfrared,schneider2021controlling}, our results have to be placed in a proper context.
In most of these previous studies, the comb repetition rate was not measured through optical detection but was picked up electrically. However, the latter signal is easily contaminated by the strong power of the injected signal from the generator due to reflections (impedance mismatch) at the QCL or weak isolation of the circulators. Thus, small differences between the laser repetition rate and the injected signal could not be detected.
On the other hand, SWIFT interferometry is an optical detection technique measuring the intermodal coherence $g$~\cite{Hillbrand-natphot-2019}, but it also fails to detect these small differences. Indeed, $g$ depends on the integrated phase noise $\phi_\mathrm{rms}$ of the repetition rate as $g=1-\phi^2_\mathrm{rms}(\tau)$ where $\tau$ is the integration time~\cite{han2020sensitivity}. Thus, a typical uncertainty of 0.1 on $g$ precludes the detection of root-mean-square phase deviations below 0.3~rad, which is much larger than the measured difference.
Therefore, our measurements and analysis shed new light on the noise dynamic of injection-locked QCL-combs.

Using the above stabilization scheme, we now demonstrate coherently-averaged DCS. Although free-running dual QCL-comb setups with interferogram computational correction were successfully used for DCS~\cite{burghoff-sciadv-2016,klocke2018singleshot,sterczewski2019computational}, these require high processing power and typically reduce the duty-cycle due to the high sampling rate. Coherent averaging bypasses these disadvantages by summing up interferograms in the temporal domain instead of spectra in the frequency domain, revealing the same results~\cite{Chen-natcomm-2018}.

\begin{figure}[t]
\centering
\includegraphics[width=\linewidth]{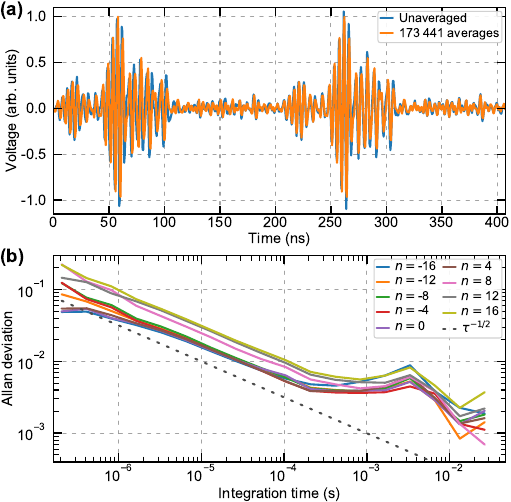}
\caption{Coherently-averaged DCS results obtained with a recording time of 71~ms. \textbf{(a)} Difference between one sample of two interferograms and an averaging of 173~441 samples. \textbf{(b)} Allan deviation of several RF comb lines.}
\label{fig:coherent_dc}       
\end{figure}

In our setup, as the repetition frequencies of the QCL-combs are RF injection locked, and as one line of the dual-comb (line 0) is stabilized, both degrees of freedom of the RF comb are then locked. The particular values we chose for $\Delta f_\text{rep} $ and for $f^\text{RF}_{0}$ allow us to obtain a free offset RF comb as $f^\text{RF}_{0} = 51 \Delta f_\text{rep} $, which is needed to avoid phase slipping between two consecutive interferograms and thus allow their averaging~\cite{Baumann-pra-2011}.
Using a digitizer, we recorded a 71~ms long temporal trace that is then processed. \autoref{fig:coherent_dc}~\textbf{(a)} shows the comparison between one sample of two interferograms and the coherent averaging of 173~441 pairs of interferograms.
The RF dual-comb spectrum obtained by Fourier transform is similar to the one displayed in \autoref{fig:DC_RF_PN}~\textbf{(a)}.

To assess the efficiency of our dual-comb setup, we computed the Allan deviation of different RF comb lines, which are presented in \autoref{fig:coherent_dc}~\textbf{(b)}.
From this result and using line $n=0$, we calculated a noise equivalent absorption of $6\times 10^{-5}$~Hz\textsuperscript{-1/2} which is similar~\cite{Gianella-oe-2020} or slightly above other dual-comb setups based on QCLs~\cite{westberg2017midinfrared,Komagata-oe-2021}.
Regarding the figure of merit~\cite{Coddington-optica-2016}, we obtained a value of $7\times 10^5$~Hz\textsuperscript{1/2} from the average of 100 lines by extrapolating the Allan deviation at 0.2~ms. This is one order of magnitude lower than the best reported value for a dual-comb setup operating in the same spectral region~\cite{Vasilyev-ol-2023}.
However, we believe that our results could be easily improved by adding a second detector for slow (>1~ms) phase and amplitude corrections~\cite{Komagata-oe-2021}, which would cancel the bump at 3~ms. Moreover, a pair of better-matched QCL-combs in terms of emission spectrum could also improve the figure of merit and hence compete with state of the art dual-comb results.

In this work, we presented a master-follower stabilization scheme for the first, to the best of our knowledge, tight-locking of two QCL-combs using a CW QCL acting as a transfer oscillator between both combs, and a high bandwidth actuator consisting of a NIR laser illuminating the front facet of a QCL~\cite{Komagata-aplphot-2023}.
The QCL-comb repetition rates were locked by RF injection using mutually phase locked signal generators to mitigate phase noise multiplication, and we performed an advanced phase noise analysis of the multi-heterodyne beating between both QCL-combs.
The results revealed that sub-radians integrated phase noise can be reached for all lines of the spectrum and also unveiled new insights on RF injection locking: we put into evidence that this stabilization technique has some limitations in the achievable noise reduction, and that this analysis can only be performed in the optical domain due to RF signal contamination.

Despite the limits, our scheme can readily be used for coherently-averaged DCS and we believe that our results will contribute towards more sensitive comb-based spectrometers in the MIR. We already mentioned potential enhancements for DCS, but improvements could also be made in the locking scheme. For instance, the stabilization of the transfer oscillator could provide a narrow-linewidth and absolute frequency reference~\cite{argence2015quantum,komagata2022absolute}, paving the way towards metrological and high resolution spectroscopic applications~\cite{Santagata-Optica-2019}.
Moreover, the line-by-line analysis method presented here is not limited to QCLs and could be implemented in other setups to characterize various comb sources, which is especially relevant in the MIR where high bandwidth and sensitive detection tools are lacking.

\begin{backmatter}
\bmsection{Funding}
Schweizerischer Nationalfonds zur Förderung der Wissenschaftlichen Forschung (40B2-1\_176584).

\bmsection{Acknowledgments} 
We thank Alpes Laser for providing the DFB-QCL used in this work. We also thank Christoph Affolderbach for lending us one generator, Mattias Beck for growing the combs, Johannes Hillbrand for mounting them, and Benoît Darquié for fruitful discussions.

\bmsection{Disclosures}
The authors declare no conflicts of interest.

\bmsection{Data availability}
Data underlying the results presented in this paper will be made available on a EUDat server.

\end{backmatter}

\bibliography{biblio}

\bibliographyfullrefs{biblio}

\end{document}